# A Survey of Energy Efficient Schemes in Ad-hoc Networks

Priya P. Patel
*Department of Computer Engineering*
*SVM Institute of technology*
*Bharuch 392-001,Gujart, India*
piyubhandari11@gmail.com

Rutvij H. Jhaveri
*Department of Computer Engineering*
*SVM Institute of Technology*
*Bharuch 392-001,Gujarat, India*
rhj_svmit@yahoo.com

***Abstract*** **- Ad hoc network is a collection of different types of nodes, which are connected in heterogeneous or homogeneous manner. It is also known as self-organizing- wireless network. The dynamic nature of ad hoc networks make them more attractive, which is used in many different applications. Every coin has two sides: one is the advantage part and other is disadvantages, in the same manner nature of ad hoc network make it more attractive from one side in other hand there are some issues too. Energy efficiency is a core factor which effects on ad hoc network in terms of battery life, throughput, overhead of messages, transmission error. For solving issues of energy constraints, different mechanisms are proposed by various researchers. In this paper, we survey various existing schemes which attempt to improve energy efficiency of different types of ad hoc routing protocol to increase network lifetime. Furthermore we outline future scope of these existing schemes which may help researches to carry out further research in this direction.**

## I. INTRODUCTION

Collection of the different nodes (i.e. PDAs, various types' sensors and laptop) in either homogeneous or heterogeneous manner is known as ad hoc network [1].These nodes have dynamic nature so that they can move in any direction randomly [1]. Self-organizing and self-configuring nature and instant response whenever application needed make an ad hoc network more reliable and attractive for today's world [2].There are various types of ad hoc networks like MANET, VANET, WSN etc. Here network refer as a heterogeneous system in which combination of tiny nodes and actuators with general purpose computing elements [3].

Ad hoc networks are very attractive for tactical communication in military and law enforcement. They are also expected to play an important role in civilian forums such as convention centres, conferences, and electronic classrooms [2]. Many other application are also there, which are monitoring air pollution, detection of forest fire, water quality observation, Disaster management, habitat monitoring, target tracking, security management, home automation and traffic control, earthquake warning and monitoring the enemy territory, agro sensing, medical field etc.

There are many aspects which effect on the ad hoc network like energy consumption, number of nodes in network, security threats, and availability of resources etc. The energy constrain is the important issues in ad hoc network.There are many terms which will effect on the energy efficiency of the network such as physical damage, limited resources, limited transmission power, limited battery power, routing overhead etc.

The rest of the paper is describe as: Section 2 describes various methods used for reducing energy consumption, Section 3 presents related work and finally, in section 4 conclusion is mentioned.

## II. METHODS TO REDUCE ENERGY CONSUMPTION

Energy management is an important issue in such networks. To increase the life of a node there are various factors like efficient battery management, transmission power management play an important role for it. Energy saving is done by two different ways: i) Power saving – to reduce power consumption ii) power control – to adjust transmission power [3].

There are various methods are developed for reducing the energy consumption of the network to increase the efficiency. There are mainly four methods [22] definedwhich are:

i) Probability based,
ii) Counterbased,
iii) Area-based and
iv) Neighbour-knowledge based methods

i) *Probability-based method:* In this method, when a node receives a broadcast message for the first time, the node rebroadcasts the message with a certain probability. If message was already received by receiver than node drop the rebroadcast message to reduce power consumption [4].

ii) *Counter-based method:*In this method, before retransmission of the message, node wait for some time periods and after that it will send the message. During this waiting time period, node maintain counter track record of all nodes. Based on this counter track node decides to drop the message. Otherwise, the node retransmits the message [4].

iii) *Area-based method:*This method is purely based on the location of the node. If the additional coverage area is less than a threshold value, all future receptions of the same message will be dropped. Otherwise, the node starts a broadcast-wait-timer [4].

iv) *Neighbour-knowledge based method:* In this method, nodes keep track of the all nodes in the network and then based on the neighbour nodes' behaviour and activities it decides whether re transmit the message or not.

## III. RELATED WORK

Energy efficiency is play an important role in the ad hoc network during routing. In recent past year, many researchers have introduced different methods and protocols related to reduce energy consumptions in different types of ad hoc network. Below shows different methods and protocols, which are used to reduce the energy consumption. We also describe these in table-1.

Santosh Kumar Das *et al.* [1] Fuzzy Based Energy Efficient Multicast Routing (FBEEMR) protocol for ad-hoc network which choose best path by which energy consumption can be reduced during routing with the help of fuzzy logic. This protocol help to increase lifetime of ad hoc network with respect to energy efficient multicast routing by calculating route lifetime values for each route.

Chong Lou *et al.* [5] proposed scheme coordinate the energy efficient routing with the sleep scheduling. In this scheme, trade-off between the reduced transmit power at senders with multi-receiver diversity and the increased receive power at forwarders with coordinated sleep scheduling. Sleep scheduling work based on 2-D grid technology and TDMA medium access control.

Tripathi *et al.* [6] proposed the cluster head selection scheme, which is used in SPIN to propagate observations which is collected from individual sensor nodes. This proposed scheme used to design program for the time interval in logical form from the start of network operation to the death of the node which is important reliable issues in WSN for many feedback applications. These proposed scheme help to resolve issues like resource limitation, energy efficiency, scalability WSN. Here C-SPIN protocol proposed which is cluster version of SPIN.

Torres *et al.* [7] introduced EBCHP – A Proactive Hierarchical routing protocol. In this protocol, the association and support strategies are used to improve performance and lifetime of the mobile ad hoc network. This proposed protocol is modified version of BCHP.

Das *et al.*[8] proposed Intelligent Energy Competency Multipath routing protocol (IECM). In proposed protocol intelligent method is used to extend lifetime by specifying a different intelligent scale to each route. The scaling phase partitioned in three phases called intelligent initiation phase, path evaluation phase and multipath route selection phase. The main goal of proposed protocol is to reduce energy consumption in WANET.

Ahyoung Lee *et al.* [9] proposed an adaptive-gossiping routing algorithm. The proposed algorithm help to saves the network bandwidth and energy consumption by reducing the duplicated routing messages. The proposed model also provide the less overhead and reduce the end to end delay.

Lim *et al.* [10] describe an adaptive VANET medium access control (MAC) layer with joint optimization for VANET (MACVS). In proposed scheme, closed loop feedback control system and joint threshold structure DP optimization scheme are used. The main goal behind this implementation is to minimize average delay and maximize packet success rate in varying nature of VANET.

Helmy *et al.* [11] introduced an optimized hierarchical routing technique. In which selection of optimal cluster heads location is based on Artificial Fish Swarm Algorithm (AFSA). Here in proposed scheme, various behaviour like preying, swarming and following of the AFSA is observed from that best one is choose by comparing them with the help of a fitness function is used.

Ruiz *et al.* [12] The Adaptive Enhanced Distance Based Broadcasting Protocol introduced. AEDB hereinafter is based on the Distance Based broadcasting protocol, and it acts differently according to local information to minimize the energy and network use, while maximizing the coverage of the broadcasting process. This protocol provide robust solution.

K.Sumathia *et al.* [13] proposed scheme named Active HELLO messaging scheme solved the major problem of MANET which is limited battery power. In this scheme nodes monitoring the link status between nodes along with the incorporation of Dynamic On Demand Routing Protocol which is used to achieve the local link connectivity information from which energy consumption can be reduced of mobile node to certain extent.

Jingmin Shi *et al.* [14] introduced a dynamic traffic regulation method which is based on virtual traffic light (VTL). In which, each vehicle can express its will- the desire of moving forward and share among another it's "will"- value and related traffic information at a traffic light controlled intersection. Based on traffic information collected in real time, the virtual light can be adaptive to the changing environment. This proposed method help to reduce waiting time which increase the energy efficiency of the network and improve the traffic efficiency.

Devasenapathy *et al.* [15] proposed energy-efficient cluster-based vehicle detection in road network using the intention numeration method (CVDRN-IN). This proposed method used to calculate energy consumption and it also reduce energy consumption using digital signature tree which reduce the amount of data transmission.

Le *et al.* [16] Coverage and Energy Strategy for wireless sensor networks is proposed. These scheme solve the major problem of the WSN which are (i) sensing coverage and (ii) network life time which is based on energy consumption. To solve first problem proposed scheme will attempt to achieve optimal coverage and energy balanced scheduling for all sensor. To solve second problem it reduce redundancy of working sensor nodes thus help to reduce power consumption. For experimental result AODV protocol is used here.

Rana *et al.* [17] Fuzzy Based Energy Efficient Multiple Cluster Head Selection Routing Protocol (FEMCHRP) proposed. This routing protocol use clustering scheme and selection of Cluster Head Nodes which contain all the details of cluster nodes. After knowing CHL, CHL send aggregated data to BS. Here which cluster head will be chosen for further operation was based on fuzzy logic. After gaining all information, when data transmission phase take place, shortest energy path used for this purpose which is done using Dijkstra Algorithm.

Layuan Li *et al.* [18] proposed an energy level based routing protocol – ELBRP, in which packet forwarding decides based on each node's energy level, this approach is core idea behind proposed protocol. This protocol not only reduce energy consumption but also increase lifetime during route. Furthermore it also helps to improve delay characteristics.

Azharuddin *et al.* [19] describe DFTR is an abbreviation of distributed energy efficient and fault tolerant routing algorithm for WSNs. The algorithm selects next-hop cluster head in energy efficient manner in the data routing phase and carefully restores the connectivity of the neighbours of a cluster head in case of its failure. The algorithm is tested extensively by considering several scenarios of WSN.

Chatterjee *et al.* [20] DSR scheme proposed which is based on ACO. The main idea behind this new scheme is – it first checks the cache for existing nodes if there is no path found then it locally broadcast the Route Request control packets to find out the routes like biological ants behave when they search for food. In addition, based on number of hops in the route pheromone will be counted. Best path which have highest pheromone count will be selected for routing. Here DSR routing protocol is used for simulation result to prove efficiency of the algorithm.

S. Kavi Priya *et al.*[21] describe heuristic technique that combine very important terms of wireless sensor networks – energy and bandwidth constraints to increase network lifetime. Here the link bandwidth is allocated based on the remaining energy making the routing solution feasible under bandwidth constraints. And for energy constraints energy efficient algorithm named nearest neighbour tree (NNT) is used.

Kiani *et al.* [22] as per title suggest, described an Efficient Intelligent Energy Routing Protocol in which intelligent routing protocol algorithm is used. The algorithm based on the reinforcement learning techniques. In proposed protocol, first of all a clustering method is applied to established network using graph. After that data transmission is done using Q-value parameter. This proposed protocol help to reach energy efficiency.

P. Rajeshwari *et al.* [23] proposed Hierarchical Energy Efficient Clustering Algorithm (HEEC). This scheme introduces a clustering algorithm, head node selection and re-electing cluster head node concept which are used to handle different energy capacity sensor nodes. HEEC is efficiently work in term of low energy consumption and balancing load on each node.

## IV. CONCLUSION

Ad hoc networks have been a popular area for researchers due to their ubiquitous nature. On one side, their self-organizing nature proves to be advantageous in many critical applications, on the other side it also raises several critical issues to be addressed such energy optimization, security empowerment, routing in resource constraint environment and many more. In this paper, we survey different mechanisms which attempt to reduce energy consumption in routing protocols for ad hoc networks.After analysingvarious schemes, we conclude that different schemes function differently in distinct situations and each has its own limitations. A scheme may prove to be more advantageous than others in one situation, but may not reach to the expectations in other situation. At the same time, it is imperative to devise an energy efficient routing scheme which works well in various scenarios for a specific type of ad hoc network.

## REFERENCES

[1] Das, Santosh Kumar, Sachin Tripathi, and A. P. Burnwal. "Fuzzy based energy efficient multicast routing for ad-hoc network." In *Computer, Communication, Control and Information Technology (C3IT), 2015 Third International Conference on*, pp. 1-5. IEEE, 2015.

[2] Pei, Guangyu, Mario Gerla, and Tsu-Wei Chen. "Fisheye state routing: A routing scheme for ad hoc wireless networks." In *Communications, 2000. ICC 2000. 2000 IEEE International Conference on*, vol. 1, pp. 70-74. IEEE, 2000.

[3] Karlof, Chris, and David Wagner. "Secure routing in wireless sensor networks: Attacks and countermeasures." *Ad hoc networks* 1, no. 2 (2003): 293-315.

[4] Kumar, Krishan, and Y. K. Jain. "Literature Survey on Energy Consumption Control for Wireless Mobile Ad-hoc Network." *IJECS* 2 (2013): 2645-2648.

[5] Lou, Chong, and Weihua Zhuang. "Energy-efficient routing over coordinated sleep scheduling in wireless ad hoc networks." *Peer-to-Peer Networking and Applications* (2015): 1-13.

[6] Tripathi, Ashutosh, Narendra Yadav, and Reena Dadhich. "Optimization of Clustering in SPIN-C and LEACH for Data Centric Wireless Sensor Networks." In *Proceedings of Fourth International Conference on Soft Computing for Problem Solving*, pp. 197-205. Springer India, 2015.

[7] Torres, Rommel, Francisco A. Sandoval, Liliana Enciso, and Samanta Cueva. "Improving Lifetime and Availability for Ad Hoc Networks to Emergency and Rescue Scenarios." In *New Contributions in Information Systems and Technologies*, pp. 979-989. Springer International Publishing, 2015.

[8] Das, Santosh Kumar, Sachin Tripathi, and A. P. Burnwal. "Intelligent Energy Competency Multipath Routing in WANET." In *Information Systems Design and Intelligent Applications*, pp. 535-543. Springer India, 2015.

[9] Lee, Ahyoung, and Ilkyeun Ra. "Network resource efficient routing in mobile ad hoc wireless networks." *Telecommunication Systems* (2015): 1-9.

[10] Lim, Joanne Mun-Yee, Yoong Choon Chang, Mohamad Yusoff Alias, and Jonathan Loo. "Joint optimization and threshold structure dynamic programming with enhanced priority scheme for adaptive VANET MAC." *Wireless Networks* (2015): 1-17.

[11] Helmy, Asmaa Osama, Shaimaa Ahmed, and Aboul Ella Hassenian. "Artificial Fish Swarm Algorithm for Energy-Efficient Routing Technique." In *Intelligent Systems' 2014*, pp. 509-519. Springer International Publishing, 2015.

[12] Ruiz, Patricia, Bernabé Dorronsoro, El-Ghazali Talbi, and Pascal Bouvry. "Finding a robust configuration for the AEDB information

dissemination protocol for mobile ad hoc networks." *Applied Soft Computing* 32 (2015): 494-508.

[13] Sumathi, K., and A. Priyadharshini. "Energy Optimization in Manets Using On-demand Routing Protocol." *Procedia Computer Science* 47 (2015): 460-470.

[14] Shi, Jingmin, Chao Peng, Qin Zhu, Pengfei Duan, Yu Bao, and Mengjun Xie. "There is a Will, There is a Way: A New Mechanism for Traffic Control Based on VTL and VANET." In *High Assurance Systems Engineering (HASE), 2015 IEEE 16th International Symposium on*, pp. 240-246. IEEE, 2015.

[15] Devasenapathy, Deepa, and Kathiravan Kannan. "An energy-efficient cluster-based vehicle detection on road network using intention numeration method." *The Scientific World Journal* 2015 (2015).

[16] Le, Nam-Tuan, and Yeong Min Jang. "Energy-Efficient Coverage Guarantees Scheduling and Routing Strategy for Wireless Sensor Networks." *International Journal of Distributed Sensor Networks* 501 (2015): 612383.

[17] Rana, Sohel, Ali Newaz Bahar, Nazrul Islam, and Johirul Islam. "Fuzzy Based Energy Efficient Multiple Cluster Head Selection Routing Protocol for Wireless Sensor Networks." (2015).

[18] Li, Layuan, Chunlin Li, and Peiyan Yuan. "An energy level based routing protocol in ad hoc networks." *Wireless Personal Communications* 81, no. 3 (2015): 981-996.

[19] Azharuddin, Md, and Prasanta K. Jana. "A distributed algorithm for energy efficient and fault tolerant routing in wireless sensor networks." *Wireless Networks* 21, no. 1 (2015): 251-267.

[20] Chatterjee, Shubhajeet, and Swagatam Das. "Ant colony optimization based enhanced dynamic source routing algorithm for mobile Ad-hoc network." *Information Sciences* 295 (2015): 67-90.

[21] Priya, S. Kavi, T. Revathi, K. Muneeswaran, and K. Vijayalakshmi. "Heuristic routing with bandwidth and energy constraints in sensor networks." *Applied Soft Computing* 29 (2015): 12-25.

[22] Kiani, Farzad, Ehsan Amiri, Mazdak Zamani, Touraj Khodadadi, and Azizah Abdul Manaf. "Efficient intelligent energy routing protocol in wireless sensor networks." *International Journal of Distributed Sensor Networks* 2015 (2015).

[23] Rajeshwari, P., B. Shanthini, and Mini Prince. "Hierarchical Energy Efficient Clustering Algorithm for WSN." (2015).

**Table -1**

| Title | Author , Publisher and year | Network types | Description | Performance Matrix | Future Scope |
|---|---|---|---|---|---|
| An Energy Level Based Routing Protocol in Ad HocNetworks | Layuan Li et al. Springer (2014) [18] | MANET | ELBRP protocol | Total consumption time, delay time, number of total death nodes, time of first node dead, | Extending proposed work for the QoS routing, energy aware multicast and any cast in mobile as hoc networks. |
| Ant colony optimization based enhanced dynamic sourcerouting algorithm for mobile Ad-hoc network | Chatterjee et al. Elsevier(2014) [20] | DTN | DSR scheme proposed based on ACO | PDR, Delay, Broken Route, Routing Overhead, Energy Consumption | Scheme can be extended to Vehicular Ad hoc Networks (VANET). |
| Finding a robust configuration for the AEDB information dissemination protocol for mobile ad hoc networks | Ruiz et al. Elsevier (2015) [12] | MANET | AEDB protocol hereinafter was based on the Distance Based broadcasting protocol | Coverage, number of forwarding, energy performance | Future Work for robust solution optimization |
| Fuzzy Based Energy Efficient Multiple ClusterHead Selection Routing Protocol for WirelessSensor Networks | Rana et al. ijcnis (2015) [17] | WSN | FEMCHRP proposed here based on clustering | Average Energy Dissipation, Average Residual Energy, lifetime | The future work can be addressed as plan to design a heterogeneous network. |
| ENERGY OPTIMIZATION IN MANETS USING ONDEMANDROUTING PROTOCOL | K.Sumathia et al. Elsevier (2015) [13] | MANET | Active HELLO messaging scheme | End to end delay and link utilization | Future work to extend lifetime and for optimal solution in MANET. |
| Energy-efficient routing over coordinated sleep schedulingin wireless ad hoc networks | Chong Lou et al. Springer (2015) [5] | Wireless ad hoc network | Energy efficient routing protocol used which was coordinate with the sleep scheduling | Packet delivery probability, Total energy consumption, Aggregate end-to-end throughput, Energy consumption per packet | Design optimal energy efficient routing protocols for wireless ad hoc network. |
| Fuzzy Based Energy Efficient Multicast Routing forAd-hoc Network | Santosh Kumar Das et al. IEEE (2015) [1] | Ad hoc network | FBEEMR protocol proposed | Route lifetime based on hop-count, Route lifetime based on energy, Route lifetime based on transmitted packet | Future work includes identify the statistic of network lifetime based on different parameters. |
| Heuristic routing with bandwidth and energy constraints in sensor networks | S. Kavi Priya et al. Elsevier(2014) [21] | WSN | Heuristic technique that combine energy and bandwidth constraints | Normalized lifetime, AVG ratio | In future, it is decided to design an algorithm that is energy efficient and bandwidth efficient by applying the fuzzy based sleep scheduling technique in order to improve the sensor network lifetime. n. |
| Energy-Efficient | Le et al. | WSN | Coverage and Energy | Network lifetime, activated | For future work, open approach |
| Coverage Guarantees Scheduling andRouting Strategy for Wireless Sensor Networks | Hindawi (2015) [16] | | Strategy for wireless sensor networks proposed | nodes, Coverage density, Coverage | on SDN be investigated to improve WSNs. |
| Optimization of Clustering in SPIN-C and LEACH for Data Centric WirelessSensor Networks | Tripathi et al. Springer (2015) [6] | WSN | Cluster head selection scheme proposed | Stability periods, number of nodes alive and PDR | Future work on dynamic replacement of node and cluster head in network after the dead and provide the better solution. |
| Improving lifetime and availability for ad hoc networks to emergency and rescue scenarios | Torres et al. Springer (2015) [7] | MANET | EBCHP protocol proposed | Battery load, Rate of sent packets, Sent packets application rate, Protocol Overhead, End to End delay, Average of Packet delay Variation | Future work on implementation of time adaptive algorithm. |
| A distributed algorithm for energy efficient and fault tolerantrouting in wireless sensor networks | Azharuddin et al. Springer(2014) [19] | WSN | DFTR algorithm proposed | Total energy and standard deviation | To design energy aware routing emphasizing partial and transient failure of the nodes and also with mobile scenario will be considered in future scope. |
| An Energy-Efficient Cluster-Based Vehicle Detection onRoad Network Using Intention Numeration Method | Devasenapathy et al. Hindawi (2015) [15] | VANET | CVDRN-IN method proposed | node energy consumption, Clustering efficiency, Node draining speed, Data aggregation count, Integrity, Traffic delivery ratio | By using more efficient and speedy encryption technique it can be possible reduce traffic delivery ratio which will be considered in future work. |
| Efficient Intelligent Energy Routing Protocol inWireless Sensor Networks | Kiani et al. Hindawi (2014) [22] | WSN | Efficient Intelligent Energy Routing Protocol named FTIEE proposed | network lifetime, Number of delivered packets, Packet delivery to network lifetime, Number of lost packets, Packet lost perdelivered packets, Delivered packet number | The intelligent routing techniques and its combining with dynamic clustering and spanning tree structures will be considered in future work. |
| Intelligent Energy Competency Multipath Routing in WANET | Das et al. Springer (2015) [8] | WANET | IECM protocol proposed | State of different route, Intelligent Scales of Different Protocol | Future work includes identify the statistic of network lifetime based on different parameters and deciding some threshold statistic for rating to each routes. |
| There Is A Will, There Is A Way--A new mechanism for traffic control based on VTL and VANET | Jingmin Shi et al. IEEE (2015) [14] | VANET | A dynamic traffic regulation method proposed based on virtual traffic light (VTL) | CO2 Emission, Vehicles Travel Time, Average in scenario time, Average waiting time | Future work considered by adding emergency vehicles to make more stable network. |
| Network resource efficient routing in mobile ad hoc wireless networks | Ahyoung Lee et al. Springer (2015) [9] | MANET | An adaptive-gossiping routing algorithm proposed | Average energy consumption, packet loss fraction, end-to-end delay, Delay × Energy, overhead, the number of data packets, throughput, average node used energy | Extending the proposed work by developing a more complex analytical model for measured energy consumption. |
| Joint optimization and threshold structure dynamic programming with enhanced priority scheme for adaptive VANET MAC | Lim et al. Springer (2015) [10] | VANET | An adaptive VANET medium access control (MAC) layer with joint optimization (MACVS) proposed | Average Delay, packet success rate | Modify MACVS algorithm with different constraint in terms of emerging nodes. |
| Hierarchical Energy Efficient Clustering Algorithm for WSN | P. Rajeshwari et al. IDOSI (2015) [23] | WSN | HEEC algorithm proposed | Delay measurement, Throughput, Source signal frequency, Destination signal frequency, Packet Drop measurement, Load balancing ratio | Useful modification in terms of decreasing packet dropping ration with respect to number of nodes increase. |
| Artificial Fish Swarm Algorithmfor Energy-Efficient Routing Technique | Helmy et al. Springer (2015) [11] | WSN | An optimized hierarchical routing technique based on AFSA proposed | Number of alive nodes for different clusters, Nodes energy residue | The future works include increasing the clusters number and the validation of proposed technique. |